\documentclass{osa-article}
\journal{oe}
\usepackage{mathtools}
\usepackage[colorinlistoftodos]{todonotes}

\begin{document}
\title{A continuous family of Exact Dispersive Quasi-Normal Modal (DQNM) Expansions for dispersive photonic structures.}

\author{Minh Duy Truong \authormark{1}, Andr\'e Nicolet \authormark{1,*}, Guillaume Dem\'esy \authormark{1}, and Fr\'ed\'eric Zolla \authormark{1}}

\address{\authormark{1} Aix Marseille Univ, CNRS, Centrale Marseille, Institut Fresnel, Marseille, France.}

\email{\authormark{*}andre.nicolet@fresnel.fr} 

\begin{abstract}
In photonics, Dispersive Quasi-Normal Modes (DQNMs) refer to optical resonant modes, solutions of spectral problems associated with Maxwell's equations for open photonic structures involving dispersive media. Since these DQNMs are the constituents determining optical responses, studying DQNM expansion formalisms is the key to model the physical properties of a considered system. In this paper, we emphasize the non-uniqueness of the expansions related to the over-completeness of the set of modes and discuss a family of DQNM expansions depending on continuous parameters that can be freely chosen. These expansions can be applied to dispersive, anisotropic, and even non-reciprocal materials. As an example, we particularly demonstrate the modal analysis on a 2-D scattering model where the permittivity of a silicon object is drawn directly from actual measurement data.
\end{abstract}

\section{Introduction}
In photonics, the interaction between the electromagnetic field (light) and matter heavily relies on the concept of resonant modes of the structure, the privileged vibrational states of the optical system \cite{lalanne,Bai13}. From a mathematical point of view, such resonant modes correspond to solutions of source-free Maxwell's equations, called Dispersive Quasi-Normal Modes (DQNMs). Under external excitation, these modes are initially loaded, then release their energy which, in turn, reflects the optical properties of the system. Therefore, it is believed that we can take advantage of these resonances (DQNMs) as building blocks in order to construct the physical characteristics of the given system. Since establishing the spectral representation of the diffracted field on a set of resonance-state basis can lead to a transparent interpretation of the numerical results, modal expansion formalisms have recently received a lot of attention \cite{Sau,Yan,vial,Mul18,Gras}.

The name Dispersive Quasi-Normal Modes (DQNMs) is derived from the fact that in practice, our media are highly dispersive and optical structures are located in unbounded media. As a result, computing these `resonant' DQNMs implies solving complicated non-Hermitian (\emph{i.e.} with complex eigenfrequencies) non-linear eigenvalue problems \cite{nonlinear}. However, numerical computations of these non-linear eigenvalue problems can still be done thanks to the current development of a powerful software library in SLEPc \cite{Hernandez}. Unfortunately, the final step to rigorously formalize modal expansions using these non-linear eigenmodes has still been a difficult mathematical task debated in recent literature \cite{lalanne,Gras}.

In an earlier paper \cite{Zolla}, we have introduced a simple approach where the exact DQNM expansion is based on the Keldysh theorem \cite{Beyn,Barel} and the concept of eigentriplets: The non-Hermitian property of dissipative systems requires the introduction of a pair of dual eigenvectors, \emph{i.e.} the `left' and `right' eigenvectors associated to the same eigenvalue. In particular, we have shown the accuracy of our modal expansion on a wide frequency range in a very strongly dispersive case where the system is closed and the permittivity is given by an artificial Lorentz model. In the present work, we report a new insight on spectral theory when applied to Maxwell's equations: There exists not just one, but a continuous family of exact DQNM expansions (in particular formulas Eq.~\eqref{elec} and Eq.~\eqref{magn}). In order to shed light on this rather strange fact, we give an example where several expansions in the family are tested for both electric and magnetic fields. Not limited to a closed system as the previous paper, we also extend our computation to open structures. Moreover, the permittivity in our calculations is no longer limited to simple models (for example the Drude-Lorentz model) but is drawn directly from actual measurement data \cite{Garcia},  which can open up many more possibilities in practical applications. The outline of the paper goes here: Section 2 introduces the general concepts of modal expansion and rational operators. It also reveals a very important spectral property of rational operators: There exists a continuous family of modal expansion formulas. Then, in section 3, we will derive the exact DQNM expansions for both electric and magnetic fields and look more closely at the non-uniqueness of these expansions. Finally, we illustrate some numerical examples in section 4 showing the effectiveness of our expansion formalism in applications with realistic materials.

\section{Modal expansion}
\subsection{Modal expansion for the Maxwell operator}
Given the domain $\Omega \subset \mathbb{R}^3$ with appropriate boundary conditions, let's consider the following Maxwell equations in harmonic regime with a time dependence in $\exp(-i\omega t)$:
\begin{align}\label{maxwell1}
\begin{split}
\nabla \times \mathbf{H}(\mathbf{r}) + i\omega \pmb{\varepsilon}(\mathbf{r},i\omega) \mathbf{E}(\mathbf{r}) & = \mathbf{J} \\
\nabla \times \mathbf{E}(\mathbf{r}) - i\omega \pmb{\mu}(\mathbf{r},i\omega) \mathbf{H}(\mathbf{r}) & = 0
\end{split}
\end{align}

In this paper, we will set $\lambda = i\omega$ to emphasize the causality of the system as a consequence of the Fourier transform: $\frac{\partial}{\partial t} \rightarrow -i\omega$ (see Subsection~\ref{sc22} for more details).

We eliminate one of the fields by combining the two first-order equations into a single second-order equation to obtain:
\begin{align}\label{maxwell2}
\mathcal{M}_{\pmb{\xi},\pmb{\chi}}(\lambda) \mathbf{u} = \mathbf{S},
\end{align}
which is expressed in terms of the Maxwell operator $\mathcal{M}_{\pmb{\xi},\pmb{\chi}}(\lambda) \coloneqq \nabla \times \left(\pmb{\xi}^{-1}(\lambda) \nabla \times \cdot \right) + \lambda^2 \pmb{\chi}(\lambda)$,
where \begin{alignat*}{2}
\mathbf{u} = \mathbf{E},\quad \pmb{\xi}=\pmb{\mu}, \quad \pmb{\chi}=\pmb{\varepsilon}, \quad & \mathbf{S}=\lambda \mathbf{J} \quad && \textrm{for electric fields}
    \\
\mathbf{u} = \mathbf{H},\quad \pmb{\xi}=\pmb{\varepsilon}, \quad \pmb{\chi}=\pmb{\mu}, \quad & \mathbf{S}=\nabla\times (\pmb{\varepsilon}^{-1} \mathbf{J}) \quad && \textrm{for magnetic fields}.
\end{alignat*}

It is worth noticing that the permittivity and permeability in the previous equations are expressed by bold notations $\pmb{\varepsilon}(\lambda)$ and $\pmb{\mu}(\lambda)$ representing `second-order tensors'. This implies the fact that there is no special restriction on the materials: they can be dispersive, anisotropic, and non-reciprocal.

The resonant modes, \emph{i.e.} DQNMs, of our electromagnetic system are indeed eigensolutions of the following spectral problem:
\begin{align}\label{eigen}
\mathcal{M}_{\pmb{\xi},\pmb{\chi}}(\lambda_n) \mathbf{v}_n = 0,
\end{align}
where $\mathbf{v}_{n}$ are the `right' eigenvectors associated to eigenvalues $\lambda_n$.

The main goal is to formalize a modal expansion equation which allows us to express the electromagnetic field $\mathbf{u}$ in terms of linear combination of $\mathbf{v}_{n}$: $\mathbf{u}=\sum_n a_n(\mathbf{S})\mathbf{v}_{n}$. This requires us to investigate the spectral properties of the operator $\mathcal{M}_{\pmb{\xi},\pmb{\chi}}(\lambda)$. In particular, we will latter point out that since the time-dispersion of $\pmb{\xi}(\lambda)$ and $\pmb{\chi}(\lambda)$, which refer to the permeability $\pmb{\mu}(\lambda)$ and  permittivity $\pmb{\varepsilon}(\lambda)$ in the case of electric fields and vice versa, can be represented by rational functions, $\mathcal{M}_{\pmb{\xi},\pmb{\chi}}(\lambda)$ can be understood as a rational operator.
\subsection{Time-dispersive materials and rational operators}
\label{sc22}
Studying the dissipation of the materials requires introducing the time-dispersion on the permittivity. This can be done by imposing a hypothetical mathematical model to the permittivity, for example, the Lorentz model \cite{alma}. However, it is more practical to assume that rational functions are a general representation of the frequency-dependence of the permittivities of media. In particular, given the polarization vector $\mathbf{P}$, the causality principle can be imposed via the constitutive relation between $\mathbf{P}$ and electric field $\mathbf{E}$, represented by the equation: $\sum_i a_i\frac{\partial^i \mathbf{P}}{\partial t^i} = \sum_j b_j\frac{\partial^{j} \mathbf{E}}{\partial t^{j}}$. Applying the Fourier transform $\frac{\partial}{\partial t} \rightarrow -i\omega$, we have $\left( \sum_i a_i (-i\omega)^i \right)\mathbf{P} = \left(\sum_j b_j (-i\omega)^j \right)\mathbf{E}$.
It implies that the electric susceptibility $\pmb{\chi}_e(i\omega)$, such that $\mathbf{P}(\omega)=\pmb{\chi}_e(i\omega)\mathbf{E}(\omega)$, must be a rational function with respect to $i\omega$ where all the coefficients are real (The $a_i$ and $b_i$ may be extended to tensors but with real coefficients); and so must the permittivity: $\pmb{\varepsilon}(i\omega)=\varepsilon_0(\mathbf{I}+\pmb{\chi}_e(i\omega))$ where $\mathbf{I}$ stands for the identity tensor.

The dispersion of permittivity can be obtained through an interpolation method \cite{Garcia} that is very accurate on a large range of frequencies and thrifty with the number of poles. The obtained rational functions are naturally causal (following Kramers-Kronig relations) and provide a natural analytic continuation of permittivities in the complex plane (including negative permittivity regions giving rise to surface plasmons).

Since the permittivity can be given a rational function, it is practically helpful to introduce the concept of rational operators: A rational operator $\mathcal{R}_L(\lambda)$ is defined as follows:
\begin{align*}
    \mathcal{R}_L(\lambda) \coloneqq \sum^N_{i=0}\mathcal{R}_i(\lambda)L_i,
\end{align*}
where $L_i$ are constant-coefficient partial differential linear operators acting on the electromagnetic field and $\mathcal{R}_i(\lambda)$ stand for rational functions with respect to $\lambda$:
\begin{align}\label{nd}
    \mathcal{R}_i (\lambda)= \dfrac{{n}_i(\lambda)}{{d}_i(\lambda)}.
\end{align}
The numerator and denominator of rational function $\mathcal{R}_i(\lambda)$ are polynomials of degree $\textrm{deg}({n}_i)$ and $\textrm{deg}({d}_i)$ respectively.

\subsection{Spectral property of rational operators}
In this subsection, we expose the spectral property of rational operators, which leads us to the formalism of modal expansion. In this paper, the following convention is used:$\langle \mathbf{w}, \mathbf{v}\rangle = \int_\Omega  \overline{\mathbf{w}(\mathbf{r})} \cdot \mathbf{v}(\mathbf{r}) d\Omega$. The `bra-ket' notation is also introduced: The `ket' vector $|\mathbf{v}\rangle $ denotes a vector such that we can simply write $\mathbf{v} = \vert\mathbf{v}\rangle$ when there is no ambiguity. On the other hand, the `bra' $\langle \mathbf{w}\vert $ represents the vector in dual space. The conjugate transpose, \emph{i.e.} Hermitian conjugate, of a bra is the corresponding ket and vice versa:$  \langle \mathbf{v}\vert^\ast =  |\mathbf{v}\rangle$. In the finite-dimensional vector space, we have that $\mathbf{v}^\ast = (\overline{\mathbf{v}})^\intercal$ where $\mathbf{v}^\intercal$ denotes the transpose and $\overline{\mathbf{v}}$ stands for the complex conjugate of $\mathbf{v}$.

Given a rational operator $\mathcal{R}_L(\lambda)$, the eigentriplets $(\lambda_n, \langle \mathbf{w}_n|, |\mathbf{v}_n\rangle)$ of $\mathcal{R}_L(\lambda)$ are defined as follows:
\begin{align}
    \langle\mathbf{w}_n| \mathcal{R}_L(\lambda_n) = 0 \qquad \text{and} \qquad
    \mathcal{R}_L (\lambda_n) |\mathbf{v}_n\rangle = 0,
    \label{eigenratio}
\end{align}
where $\langle \mathbf{w}_n|$ and $|\mathbf{v}_n\rangle$ are the `left' and `right' eigenvectors corresponding to the same eigenvalue $\lambda_n$. We assume that the operator $\mathcal{R}_L(\lambda)$ is diagonalizable and all the eigenvalues $\lambda_n$ are simple. This requires $\mathcal{R}'_L(\lambda) \neq 0$ where $\mathcal{R}^\prime_L(\lambda) \coloneqq \dfrac{d \mathcal{R}_L(\lambda)}{d\lambda}$ is the complex derivative of $\mathcal{R}_L(\lambda)$ with respect to the complex variable $\lambda$. In practice, the complex derivative of operators is obtained by taking the complex derivative of coefficients (functions of $\lambda\in\mathbb{C}$) in the operator, \emph{i.e.} the rational functions $\mathcal{R}_i$ in our case.

The rational operator can be multiplied by a $N_D$-degree polynomial $D(\lambda)$ that is divisible by all the polynomials in the denominators to get the $N_N$-degree polynomial operator $\mathcal{N}_L(\lambda)=\mathcal{R}_L(\lambda)D(\lambda)$. The degrees $N_N$ and $N_D$ are computed by the following equalities: 
\begin{align}\label{d+n}
    N_N  = \sup_i \left(\deg(n_i)+\sum_{h\neq i}\deg(d_h) \right) \quad \text{and} \quad
    N_D  = \sum_{i}\deg(d_i),
\end{align}
if $\textrm{root}(d_i) \neq \textrm{root}(d_j)$  for $\forall i, j \in \{0,1,\ldots, N\}$. 

Through the process of linearization, the polynomial operator $\mathcal{N}_L(\lambda)$ can be expressed in terms of a system of linear problems. From there, we have proved that the solution $\mathbf{u}$ of the non-homogeneous problem $\mathcal{R}_L(\lambda) \mathbf{u} = \mathbf{S}$, which is also the solution of $\mathcal{N}_L(\lambda) \mathbf{u} = D(\lambda)\mathbf{S}$, can be expanded in the form of the following quasi-normal modal expansion formulas:
\begin{equation}\label{PQNM}
\mathbf{u} = \sum_n  \dfrac{g_\sigma(\lambda_n)}{g_\sigma(\lambda)} \dfrac{1}{\lambda - \lambda_n} \dfrac{\langle\mathbf{w}_n, D(\lambda)\mathbf{S}\rangle}{\langle\mathbf{w}_n,\mathcal{N}'_L(\lambda_n) \mathbf{v}_n\rangle} \,\mathbf{v}_n
\end{equation}
where $g_\sigma(\lambda)$ is an arbitrary polynomial of degree $\sigma$ with $\sigma \in \lbrace 0, 1, \ldots, N_N-1 \rbrace $.

By replacing $\mathcal{N}_L'(\lambda_n)\mathbf{v}_n=D(\lambda_n)\mathcal{R}_L'(\lambda_n)\mathbf{v}_n$ (using Eq.~\eqref{eigenratio}), we obtain:
\begin{equation}\label{RQNM}
\mathbf{u} = \sum_n  \dfrac{g_\sigma(\lambda_n)D(\lambda)}{g_\sigma(\lambda)D(\lambda_n)} \dfrac{1}{\lambda - \lambda_n} \dfrac{\langle\mathbf{w}_n, \mathbf{S}\rangle}{\langle\mathbf{w}_n,\mathcal{R}'_L(\lambda_n) \mathbf{v}_n\rangle} \,\mathbf{v}_n = \sum_n  \dfrac{f_\rho(\lambda_n)}{f_\rho(\lambda)} \dfrac{1}{\lambda - \lambda_n} \dfrac{\langle\mathbf{w}_n, \mathbf{S}\rangle}{\langle\mathbf{w}_n,\mathcal{R}'_L(\lambda_n) \mathbf{v}_n\rangle} \,\mathbf{v}_n
\end{equation}
where $f_\rho(\lambda)$ is an arbitrary polynomial of degree $\rho$ with $\rho \in \lbrace 0, 1, \ldots, N_N-N_D-1 \rbrace $, and  $\mathcal{R}^\prime_L(\lambda_n)=\mathcal{R}^\prime_L(\lambda)\vert_{\lambda = \lambda_n}$. It turns out that if $N_N-N_D<1$, the family of expansions with the $f_\rho(\lambda)$ in Eq.~\eqref{RQNM} is no longer valid.

\section{The exact DQNM expansions for electromagnetic problem}
Since $\mathcal{M}_{\pmb{\xi},\pmb{\chi}}(\lambda)$ can be seen as a rational operator, it is possible to derive the DQNM expansion for an electromagnetic problem with dispersive media using the expansion formula Eq.~\eqref{RQNM}. In order to do that, we firstly have to clarify the meaning of the `left' eigenvalue problem of  $\mathcal{M}_{\pmb{\xi},\pmb{\chi}}(\lambda)$: $\langle\mathbf{w}_n\vert \mathcal{M}_{\pmb{\xi},\pmb{\chi}}(\lambda_n)=0$. For a deeper discussion of the `left' eigenvalue problem and adjoint operator, we refer the reader to the appendix.

\subsection{The exact DQNM expansions for electric fields}
Equipped with the eigentriplets of the operator $\mathcal{M}_{\pmb{\xi},\pmb{\chi}}(\lambda)$, our next step is to apply Eq.~\eqref{RQNM} in order to obtain the DQNM expansion for electric fields. Given the operator $\mathcal{M}_{\pmb{\mu},\pmb{\varepsilon}}$ with the eigentriplets $(\lambda_n, \langle \mathbf{E}_{ln}\vert,  |\mathbf{E}_{rn}\rangle)$, the modal expansion of the solution $\mathbf{E}$ of wave equation $\mathcal{M}_{\pmb{\mu},\pmb{\varepsilon}} \mathbf{E} =\mathbf{S}$ appears to be:
\begin{align}\label{elec}
    \mathbf{E} = \sum_n  \dfrac{f_\rho(\lambda_n)}{f_\rho(\lambda)} \dfrac{1}{\lambda - \lambda_n} \dfrac{\langle\mathbf{E}_{ln}, \mathbf{S}\rangle}{\left\langle\mathbf{E}_{ln},\mathcal{M}'_{\pmb{\mu},\pmb{\varepsilon}}(\lambda_n) \mathbf{E}_{rn}\right\rangle} \,\mathbf{E}_{rn},
\end{align}
where the inner product in the denominator can be computed explicitly as follows:
\begin{align*}
    \langle\mathbf{E}_{ln} ,\mathcal{M}'_{\pmb{\mu},\pmb{\varepsilon}}(\lambda_n)\mathbf{E}_{rn}\rangle =  \int_\Omega \left[ \overline{\mathbf{E}_{ln}}\cdot\left( \nabla\times((\pmb{\mu}^{-1}(\lambda_n))^\prime\nabla\times\mathbf{E}_{rn})\right) + \overline{\mathbf{E}_{ln}} \cdot \left( (\lambda_n^2\pmb{\varepsilon}(\lambda_n))'\mathbf{E}_{rn} \right) \right]\,d\Omega.
\end{align*}
where the prime denotes the complex derivative of functions: $f'(\lambda_n) \coloneqq \left. \frac{d f(\lambda)}{d\lambda} \right\vert_{\lambda=\lambda_n}$.

The final step is to identify the value of degree $\rho$ in Eq.~\eqref{elec} by finding out the value of $N_D$ and $N_N$. Since the frequency-dependence of permittivity can be efficiently represented by a rational function \cite{Garcia}, it is reasonable to consider that the orders of polynomials in the numerator and denominator of such rational functions are equal. Indeed, the permittivity is given by $\pmb{\varepsilon} = \varepsilon_0 (\mathbf{I}+ \pmb{\chi}_e)$ where $\varepsilon_0$ is the electric permittivity of free space. And at very high frequencies, the susceptibility $\pmb{\chi}_e$ decreases asymptotically as $\pmb{\chi}_e \propto  (1/\omega^2) \mathbf{I}$ and $\pmb{\varepsilon} \rightarrow \varepsilon_0\mathbf{I}$. Thus, we can assume that  $\pmb{\varepsilon}$ is given by rational functions whose numerator and denominator are polynomials of the same degree. Then, with the same argument for permeability, Eq.~\eqref{d+n} implies that
\begin{align}\label{nnnd}
    N_N-N_D=2.
\end{align}
In order to understand the reason behind Eq.~\eqref{nnnd}, let's assume our media are isotropic. Then, our Maxwell operator can be rewritten as follows:
\begin{align}
    \mathcal{M}_{\pmb{\mu},\pmb{\varepsilon}}(\lambda) = \underbrace{\mu^{-1}(\lambda)}_{n_0/d_0} \underbrace{\nabla\times\left(\nabla \times \cdot \right)}_{L_0} + \underbrace{\lambda^2 \varepsilon(\lambda)}_{n_1/d_1} \underbrace{\mathbf{I}}_{L_1},
\end{align}
where $\mathbf{I}$ stands for identity operator.
It is easy to recognize $\mathcal{M}_{\pmb{\mu},\pmb{\varepsilon}}(\lambda)$ in the previous equation as a rational operator with $\textrm{deg}(n_0)= \textrm{deg}(d_0)$ and $\textrm{deg}(n_1)=\textrm{deg}(d_1)+2$. Using Eq.~\eqref{d+n}, it is easy to verify that $N_N-N_D=2$. It is worth pointing out that Eq.~\eqref{nnnd} holds for all dispersive, anisotropic and even non-reciprocal materials.

As a result, $f_\rho(\lambda)$ in Eq.~\eqref{elec} can be any arbitrary polynomial up to degree 1. This means $f_\rho(\lambda)$ would take the form $f_\rho(\lambda) = \alpha+\lambda \beta$ with $\forall \alpha, \beta \in \mathbb{C}$ and  $\vert\alpha\vert+\vert\beta\vert \neq 0$. As a consequence, there exists a continuous family of  expansion formulas for the electric field $\mathbf{E}$ of the operator $\mathcal{M}_{\pmb{\mu},\pmb{\varepsilon}}(\lambda) = \nabla \times \left(\pmb{\mu}^{-1}(\lambda) \nabla \times \cdot \right) + \lambda^2 \pmb{\varepsilon}(\lambda)$. For example, if we set $f_\rho(\lambda)=1$, the modal expansion is similar to the formalization established using the Keldysh theorem in our earlier paper \cite{Zolla}. On the other hand, with the choice $f_\rho(\lambda)=\lambda$, Eq.~\eqref{elec} becomes:
\begin{align}\label{elec2}
\mathbf{E} = \sum_n \dfrac{\lambda_n}{\lambda(\lambda - \lambda_n)} \dfrac{\int_\Omega \overline{\mathbf{E}_{ln}}\cdot \mathbf{S}\,d\Omega}{\int_\Omega \left[ \overline{\mathbf{E}_{ln}}\cdot\left( \nabla\times((\pmb{\mu}^{-1}(\lambda_n))^\prime\nabla\times\mathbf{E}_{rn})\right) + \overline{\mathbf{E}_{ln}} \cdot \left( (\lambda_n^2\pmb{\varepsilon}(\lambda_n))'\mathbf{E}_{rn} \right) \right]\,d\Omega} \,\mathbf{E}_{rn},
\end{align}
which indeed recovers other expansion formulas previously proposed in the literature \cite{lalanne,Sau}.
As an example, we recall equations (4) and (7) from reference \cite{Sau} using our new notation:
\begin{align}\label{eq:Sau}
\begin{split}
    \mathbf{E}(\mathbf{r},\omega)& \approx \sum_n\alpha_n(\omega)\mathbf{E}_{rn}(\mathbf{r})\\
\alpha_n(\omega) &= \dfrac{-\omega \mathbf{p}\cdot \mathbf{E}_{rn}(\mathbf{r}_0)}{(\omega-\tilde{\omega}_n)\int_\Omega \left[\mathbf{E}_{rn} \cdot \frac{\partial(\omega \varepsilon)}{\partial\omega} \mathbf{E}_{rn}- \mathbf{H}_{rn} \cdot \frac{\partial(\omega \mu)}{\partial\omega} \mathbf{H}_{rn} \right]\,d\Omega} + f_n(\omega)
\end{split}
\end{align}
where the source $\mathbf{S}$ is a dipole located at the point $\mathbf{r}_0$ and $f_n(\omega)$ is a nonresonant background that is negligible, according to \cite{Sau}.
In the case where the media are reciprocal, \emph{i.e.} the `left' eigenvectors are the complex conjugate of their `right' counterparts (as proved in the appendix), we can prove that Eq.~\eqref{elec2} implies Eq.~\eqref{eq:Sau} considering $\lambda=i\omega$, $\mathbf{S}=\lambda \mathbf{J}$, and $\mathbf{J}=-i\omega\mathbf{p}\delta(\mathbf{r}-\mathbf{r}_0)$. 

\subsection{The exact DQNM expansions for magnetic fields}
Similarly as in the case of electric fields, a DQNM expansion can be obtained for magnetic fields. Given the operator $\mathcal{M}_{\pmb{\varepsilon},\pmb{\mu}}$ with the eigentriplets $(\lambda_n, \langle \mathbf{H}_{ln}\vert,  |\mathbf{H}_{rn}\rangle)$, the modal expansion of the solution $\mathbf{H}$ of wave equation $\mathcal{M}_{\pmb{\varepsilon},\pmb{\mu}} \mathbf{H} =\mathbf{S}$ appears to be:
\begin{align}\label{magn}
    \mathbf{H} = \sum_n  \dfrac{f_\rho(\lambda_n)}{f_\rho(\lambda)} \dfrac{1}{\lambda - \lambda_n} \dfrac{\langle\mathbf{H}_{ln}, \mathbf{S}\rangle}{\left\langle\mathbf{H}_{ln},\mathcal{M}'_{\pmb{\varepsilon},\pmb{\mu}}(\lambda_n) \mathbf{H}_{rn}\right\rangle} \,\mathbf{H}_{rn},
\end{align}
where the inner product in the denominator can be computed explicitly as follows:
\begin{align*}
    \langle\mathbf{H}_{ln} & ,\mathcal{M}'_{\pmb{\varepsilon},\pmb{\mu}}(\lambda_n)\mathbf{H}_{rn}\rangle =  \int_\Omega \left[ \overline{\mathbf{H}_{ln}}\cdot\left(\nabla\times( (\pmb{\varepsilon}^{-1}(\lambda_n))^\prime\nabla\times\mathbf{H}_{rn})\right) + \overline{\mathbf{H}_{ln}} \cdot \left( (\lambda_n^2\pmb{\mu}(\lambda_n))'\mathbf{H}_{rn} \right) \right]\,d\Omega.
\end{align*}

Following the same argument as the previous subsection,  $f_\rho(\lambda)$ in Eq.~\eqref{magn} can be any arbitrary polynomial up to degree 1. This means $f_\rho(\lambda)$ would take the form $f_\rho(\lambda) = \alpha+\lambda \beta$ with $\forall \alpha, \beta \in \mathbb{C}.$

\subsection{A continuous family of exact DQNM expansions for electromagnetic wave}
\label{sec34}
Thus far, we have already mentioned the existence of a continuous family of modal expansions for electromagnetic fields, for example, Eq.~\eqref{elec} for electric fields. This property results from the over-completeness of the set of eigensolutions of non-linear in frequency operators in general and Maxwell operators in particular.  Still, it is easy to emphasize
the counter-intuitiveness of the non-uniqueness of DQNM expansion. For example, it is possible to discard the contribution of any resonant mode from the DQNM expansion of the Maxwell operator.

In order to fully understand the previous property, consider the DQNM expansion for electric fields Eq.~\eqref{elec} with $f_\rho(\lambda)=\alpha + \lambda \beta$, with $\forall \alpha,\beta \in \mathbb{C}$ and $\vert\alpha\vert+\vert\beta\vert \neq 0$. Then, among $n$ eigenvalues $\lambda_n$, let's choose specifically an eigenvalue  $\tilde{\lambda}$ with a corresponding eivenvector $\tilde{\mathbf{E}}$. The issue arises if we choose the function $f_\rho(\lambda)$ such that $\alpha = -\tilde{\lambda} \beta$. Then the factor $f_\rho(\lambda_n)$ will become $0$ when $\lambda_n=\tilde{\lambda}$. This implies that at the eigenvalue $\tilde{\lambda}$, the resonant mode, \emph{i.e.} the eigenvector $\tilde{\mathbf{E}}$ will be excluded in the expansion of the electric field $\mathbf{E}$. \textit{This is an extremely unexpected spectral property which has never been discovered by any modal expansion formalism in the literature!}   

Moreover, the non-uniqueness of DQNM expansions also raises a question of which formula of $f_\rho(\lambda)$ we should choose to perform the expansion. We notice that the expansion for electric fields Eq.~\eqref{elec} will explode at the rate $\frac{1}{\lambda-\lambda_n}$ when $\lambda$ moves close to $\lambda_n$. In practice, this singularity never occurs because $\lambda_n$ are complex resonant frequencies while $\lambda$ only accepts imaginary values. Unfortunately, the problem will raise if we choose, for example, $f_\rho(\lambda)=\lambda-\tilde{\lambda}_1$ where $\tilde{\lambda}_1$ is in the vicinity of one of resonant modes $\lambda_n$. Then the expansion Eq.~\eqref{elec} will explode at the rate $\frac{1}{(\lambda-\tilde{\lambda}_1)^2}$ when $\lambda$ moves close to $\tilde{\lambda}_1$. In fact, we will numerically demonstrate in the next section that the expansion Eq.~\eqref{elec} is no longer accurate in the vicinity of one of the resonant modes $\lambda_n$ if $f_\rho(\lambda)=\lambda-\tilde{\lambda}_1$. 

\section{Numerical examples}
\begin{figure}[htbp] 
\centering\includegraphics[width=.55\linewidth]{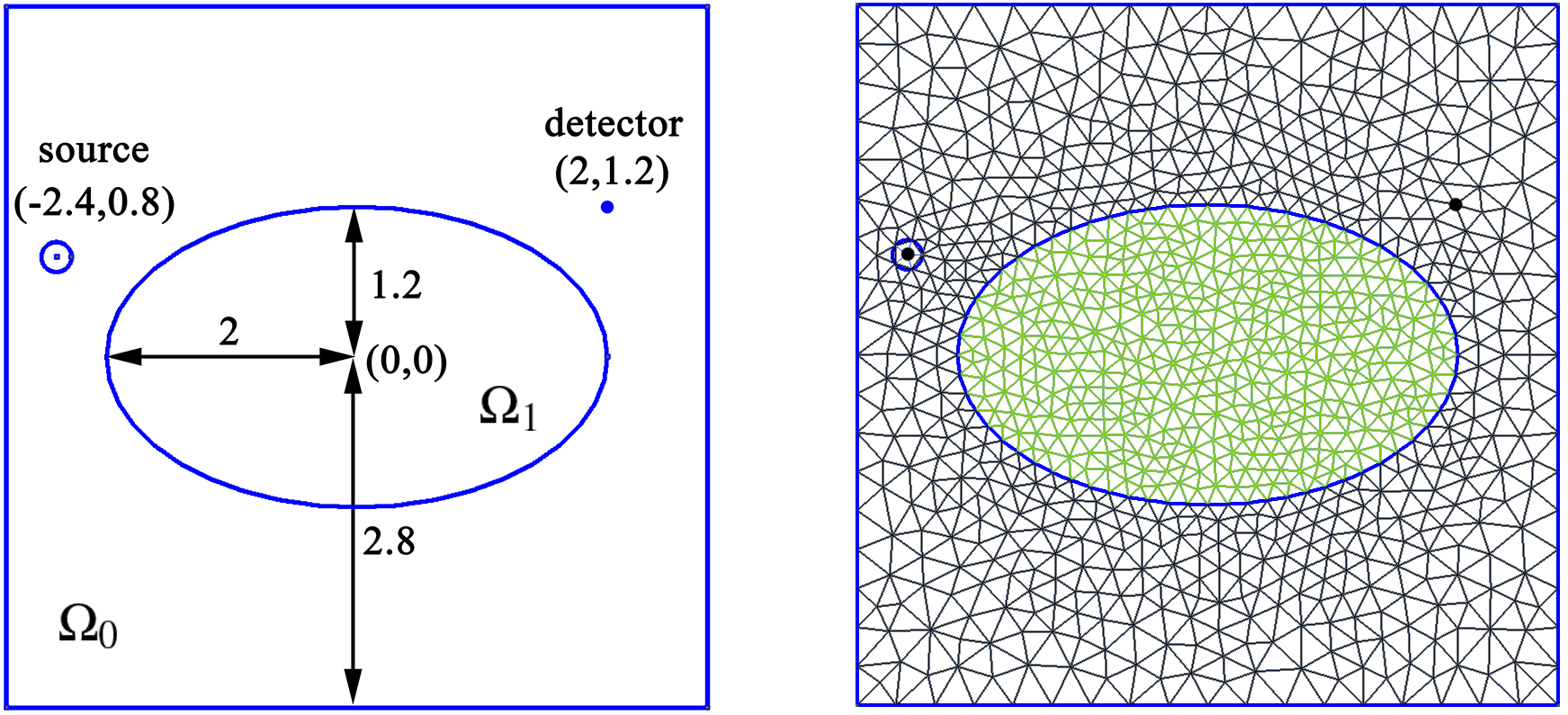}
\caption{(Left) A 2-D square box containing an elliptical scatterer $\Omega_1$: The major and minor radius of the ellipse are 2 and 1.2 respectively. The length of the side of a square is 5.6 (All lengths are measured in $(\times 10^{-1} \mu m)$). (Right) The Finite Element Mesh.}
\label{fig:geo}
\end{figure}

We will illustrate numerical results in the geometry of an object shaped like an ellipse $\Omega_1$ inside a perfectly conducting vacuum square $\Omega_0$ (see Fig.~\ref{fig:geo}). The parameters of the structure are chosen in such a way
that the material and geometric resonances highly interact with each other. In particular, we want to exhibit the problem where the geometric resonances are in the vicinity of the electric poles of the permittivity.

The elliptic scatterer is made of silicon whose dispersive permittivity is given by the multi-pole `Lorentz' model:
\begin{align}\label{pole}
    \varepsilon_{\textrm{Si}}(\mathbf{r},\omega) = 1 + \sum_{i=1}^{N_p} \left( \dfrac{A_i(\mathbf{r})}{\omega - \omega^\varepsilon_i} - \dfrac{\overline{A_i(\mathbf{r})}}{\omega + \overline{\omega^\varepsilon_i}}  \right),
\end{align}
where $N_p$ is the total number of poles in our model and $\omega^\varepsilon_i$ stands for the complex values of electric poles. For instance, the data of silicon can be extracted from the following table (The frequency unit is $\times 10^{14} (\textrm{rad}.s^{-1})$):
\begin{center}
 \resizebox{12cm}{!}{
 \begin{tabular}{|| c | c c c c||} 
 \hline\textbf{}
 $i$ & 1 & 2 & 3 & 4 \\ [0.5ex] 
  \hline
 $A_i(\textbf{r})$ & $-165.959-20.199i$ & $-113.424+89.872i$  & $-41.362+41.091i$ & $-34.218-47.163i$  \\ 
 \hline
 $\omega^\varepsilon_i$ & $64.605-4.127i$ & $72.079-14.16i$ & $51.186-2.109i$ & $59.553-4.219i$  \\ 
 \hline
\end{tabular}}
\end{center}

\begin{figure}[htbp] 
\centering\includegraphics[width=.56\linewidth]{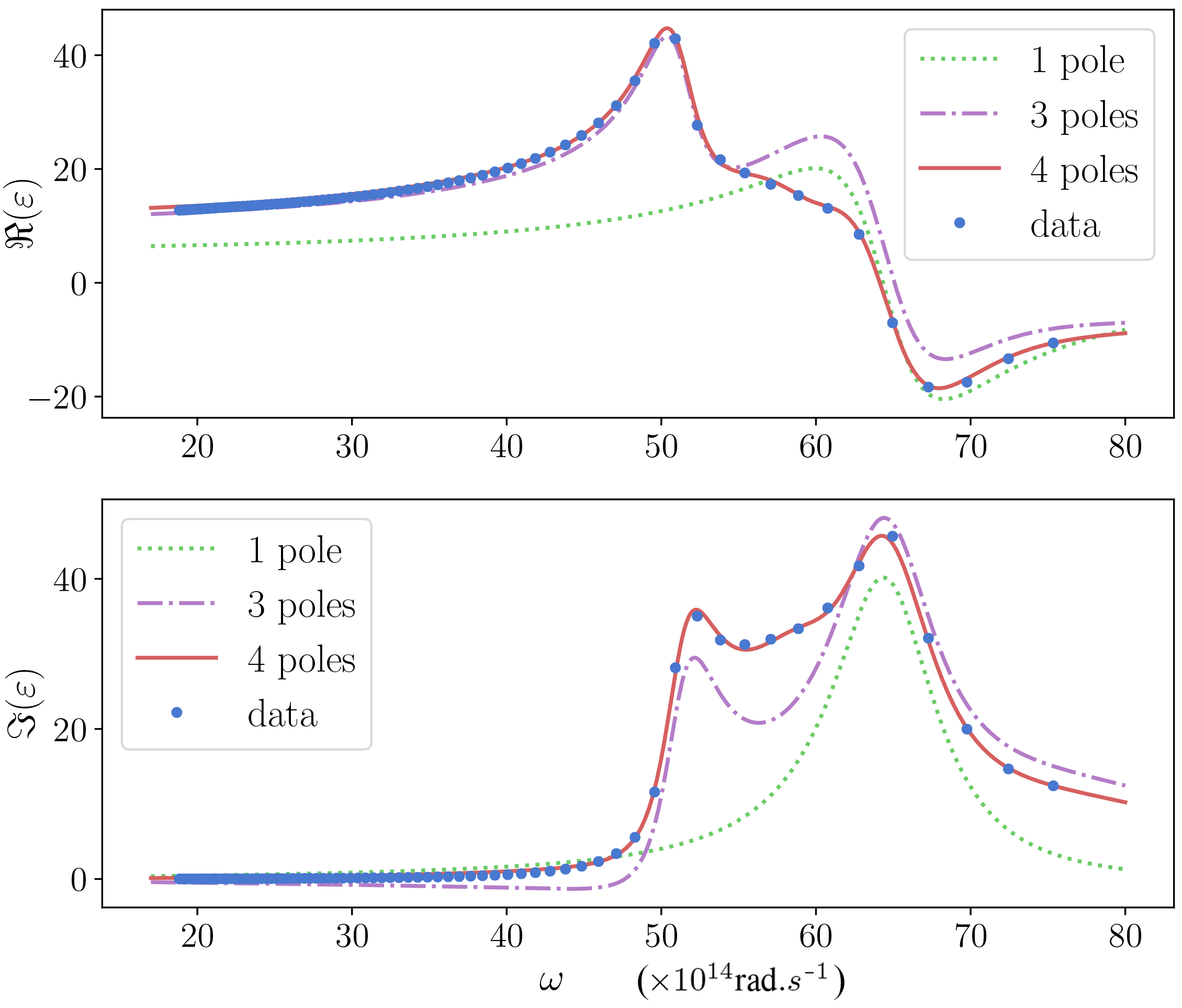}
\caption{The real part (top) and imaginary part (bottom) of the permittivity of silicon computed with different numbers of poles. The blue dots represent the actual measurement data of silicon.}
\label{fig:si}
\end{figure}

In fact, by adding more poles to Eq.~\eqref{pole} up to $N_p=4$, we can draw an exact formula of permittivity for any realistic materials from measurement data \cite{Garcia,Green1995}. For example, we plot in Fig.~\ref{fig:si} the real and imaginary part of the permittivity of silicon calculated based on Eq.~\eqref{pole} using different numbers of poles. According to Fig.~\ref{fig:si}, the more number of poles, the better resembling realistic data our model of permittivity. At the same time, the relative permittivity of vacuum is set to be constant $\varepsilon_{\textrm{vac}} = 1$. Then, the whole structure will be illuminated by the Dirac delta source $\mathbf{S}=\delta(\mathbf{r}_S)$ whose coordinates is given by $\mathbf{r}_S =(-2.4,0.8) (\times 10^{-1} \mu m)$, see Fig.~\ref{fig:geo}. The maximum element size is set to be $0.03 \, \mu m$, in comparison to the smallest wavelength  in vacuum $0.2355 \, \mu m$ (equivalent to the highest frequency of the spectrum).

\subsection{A family of DQNM expansion}
The aim of this subsection is to exemplify the non-uniqueness of our DQNM expansion for the geometry in Fig.~\ref{fig:geo}. In particular, we want to compare the results obtained by solving the direct electric scattering problem $\mathcal{M}_{\pmb{\mu},\pmb{\varepsilon}} \mathbf{E} =\mathbf{S}$ in TE polarization and the field $\mathbf{E}$ constructed using DQNM expansion Eq.~\eqref{elec} with different functions $f_\rho(\lambda)$. For the sake of clarification, let's begin with a 1-pole model of permittivity, \emph{i.e.} $N_p=1$.  
The $3$-D electrodynamic eigenvalue computations require genuine edge elements to avoid spurious modes but, in our $2$-D case, we use longitudinal fields $E_z \mathbf{e}_z$ or $H_z \mathbf{e}_z$ and the associated edge elements reduce to the Lagrange basis elements, here second-order, for the (scalar) component $E_z$ or $H_z$.

The eigenvalue problem is numerically solved thanks to recent versions of SLEPc library \cite{Hernandez,demsy}  available in the Finite Element Method (FEM) open source package ONELAB/Gmsh/GetDP \cite{getdp1,getdp2} that we are using to implement our models.

\begin{figure}[htbp] 
\centering\includegraphics[width=0.72\linewidth]{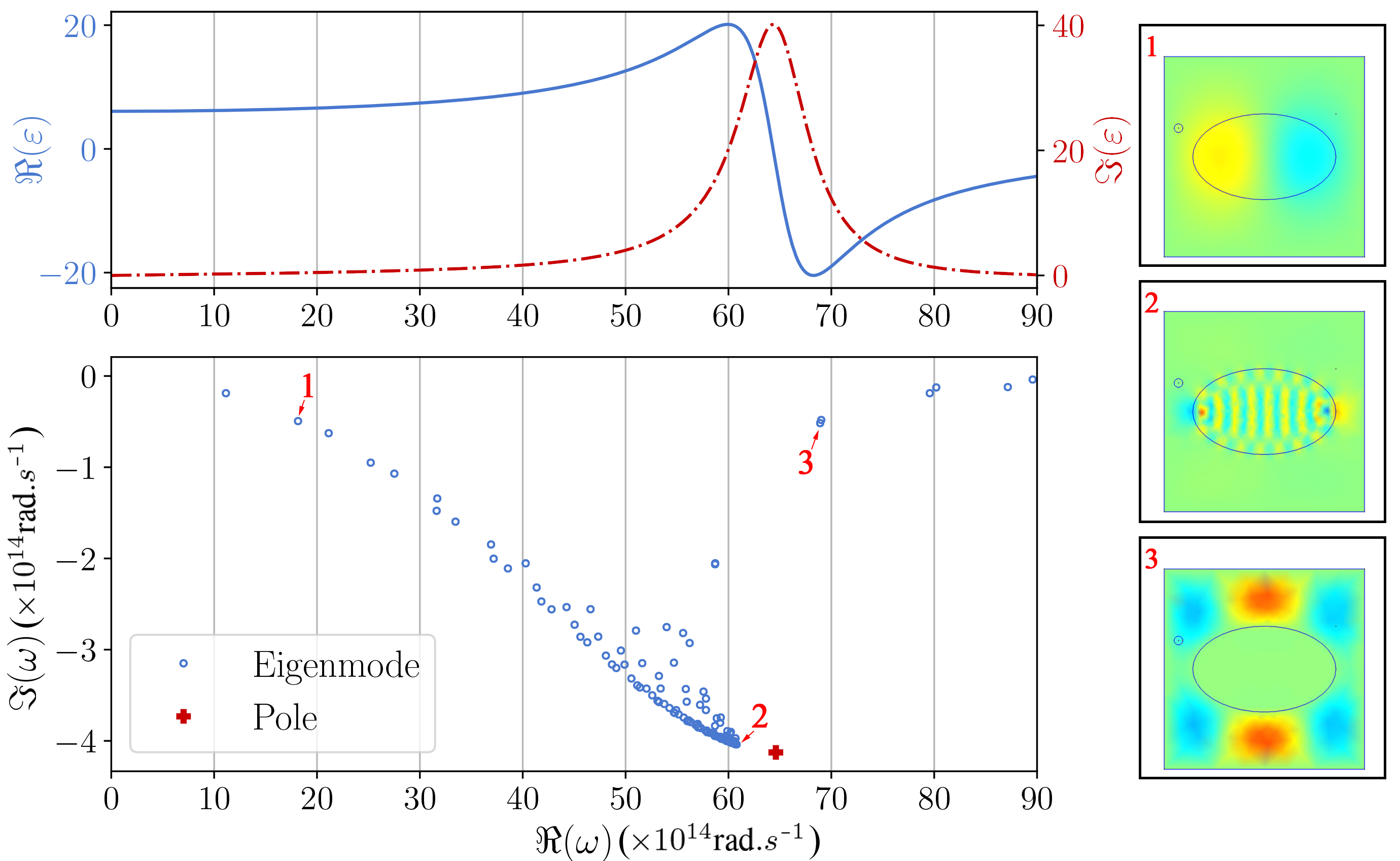}
\caption{Spectrum of complex eigenfrequencies (bottom left) corresponding to the 1-pole relative permittivity (top left). Three eigenfields (real part) are depicted at the right (the blue color of the field maps
indicates the minimum value and the red is the maximum).}
\label{fig:eigen1}
\end{figure}

The complex resonances are shown in Fig.~\ref{fig:eigen1}. We emphasize the existence of an accumulation point in the vicinity of the electric pole $\omega^\varepsilon_1$ (red cross) where $\varepsilon_\textrm{Si} \rightarrow \infty$. The modes around the pole concentrate inside the scatterer and have spatial frequency tending to infinity (for instance mode 2 in Fig.~\ref{fig:eigen1}), which distinguishes them from conventional modes with
eigenfield located in the
free-space background (see mode 1 and mode 3 in Fig.~\ref{fig:eigen1}). It is worth reminding that in the case of closed structures, the spectrum of eigenfrequencies is indeed symmetric through the imaginary axis. Since the contribution of the modes on the left half of the complex plane is numerically insignificant (the factor $\frac{1}{\lambda-\lambda_n}$ of these modes is relatively small compared to their counterparts on the right half of the complex plane), in this paper, we do not include them.

\begin{figure}[htbp] 
\centering\includegraphics[width=1\linewidth]{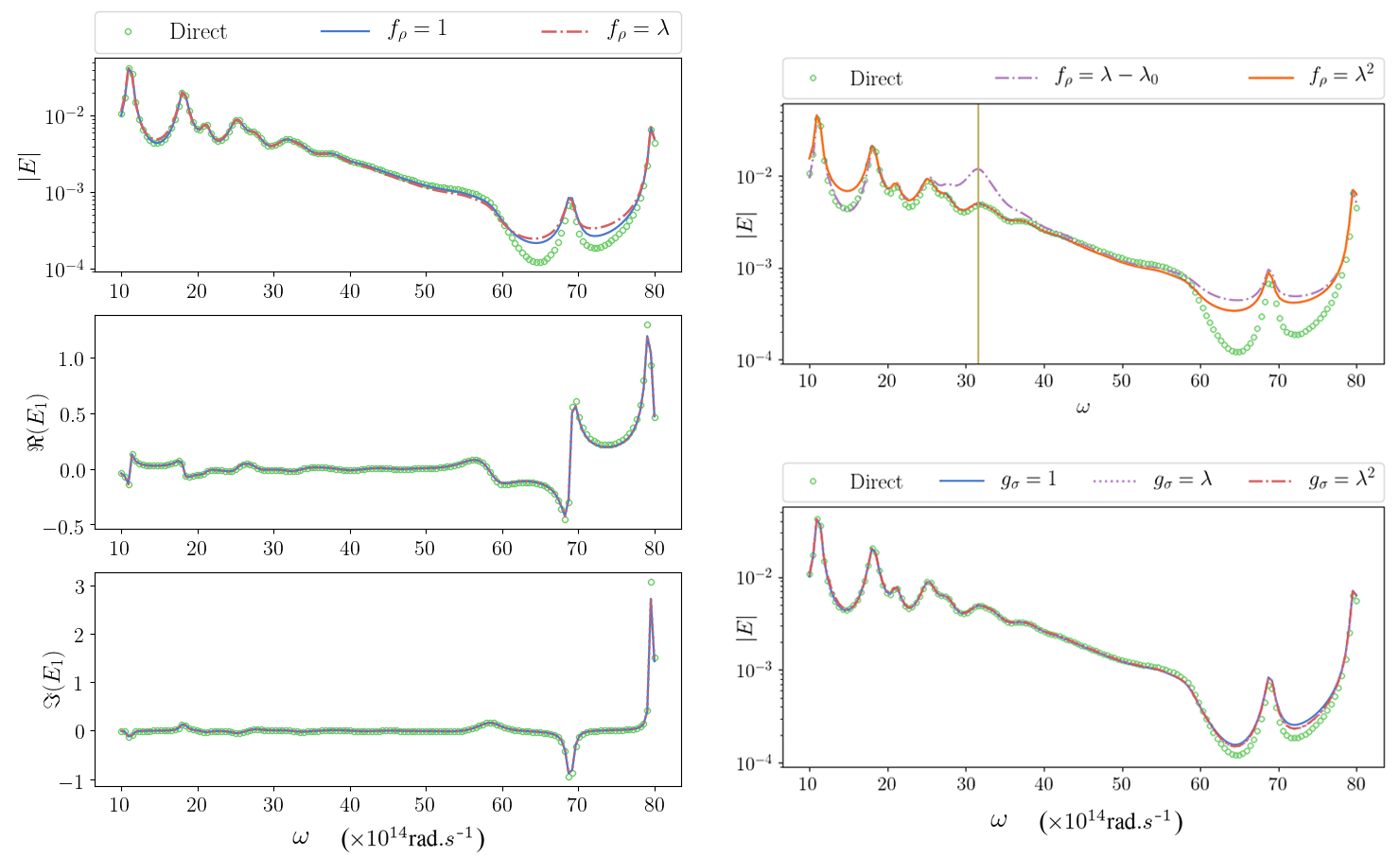}
\caption{On the left: Scattered field $\mathbf{E}$ obtained by different expansion formulas Eq.~\eqref{elec} or by solving a direct problem classically (green dots) corresponding to the 1-pole permittivity. (Top-left) Integral over $\Omega_1$ of the norm of the electric field $\int_{\Omega_1}\vert \mathbf{E}\vert d\Omega$. (Middle-left and bottom-left) The real and imaginary part of the electric field calculated at the detector point. \\
On the right: (Top-right) Scattered field $\mathbf{E}$ reconstructed by  Eq.~\eqref{elec} with $f_\rho(\lambda)=\lambda-\lambda_0$  and $f_\rho(\lambda)=\lambda^2$. The orange vertical line indicates the value $\Re(\omega_0)$. (Bottom-right) Scattered field $\mathbf{E}$ rebuilt by Eq.~\eqref{PQNM}.}
\label{fig:e1}
\end{figure}

The numerical comparison between the direct computation of electric field $E$ and the reconstruction based on the DQNM expansion Eq.~\eqref{elec} where $f_\rho(\lambda)=1$, and $f_\rho(\lambda)=\lambda$ is displayed in 3 subfigures on the left of Fig.~\ref{fig:e1}. The results are represented in three figures corresponding to three quantities: the norm of
electric field in the domain $\Omega_1$: $\int_{\Omega_1} \vert \mathbf{E} \vert d\Omega$ (displayed in logscale at the top-left subfigure of Fig.~\ref{fig:e1}); the real and imaginary part of the electric field $\mathbf{E}_1$ calculated at the detector point  in Fig.~\ref{fig:geo} (represented by the middle-left and bottom-left subfigures of Fig.~\ref{fig:e1}). The DQNM expansion in the cases where $f_\rho(\lambda)=1$ (blue lines) and $f_\rho(\lambda)=\lambda$ (red lines) demonstrates excellent agreement with green dots obtained by directly solving the scattering problem except in the vicinity of $\Re(\omega^\varepsilon_1)$. 

In addition, the norm of the field $\int_{\Omega_1} \vert \mathbf{E} \vert d\Omega$ reconstructed using the DQNM expansion formula with $f_\rho(\lambda)=\lambda-\lambda_0$ and $f_\rho(\lambda)=\lambda^2$ are shown in the top-right subfigures of Fig.~\ref{fig:e1}.
The value  $\lambda_0 = i\omega_0$ is selected such that   $\omega_0=31.628-1.478i$ $(\times 10^{14} \textrm{rad}.s^{-1})$, which is almost coincide with an eigenfrequency. With $f_\rho(\lambda)=\lambda-\lambda_0$ (purple lines), we notice discrepancies around the frequency $\Re(\omega_0)$, which confirms our prediction in subsection \ref{sec34} about the singularity at the roots of $f_\rho(\lambda)$. As a result, it is recommended to choose $f_\rho(\lambda)=\alpha + \lambda \beta$ such that the value $-\alpha/\beta$ is far away from our domain of interest. Unsurprisingly, when the degree of the polynomial  $f_\rho(\lambda)$ is higher than 1, \emph{i.e.} $f_\rho(\lambda)=\lambda^2$ (orange line), we see the less accurate in the numerical results of the DQNM expansion since it is not an appropriate formula.

Finally. it is worth noting that if we choose to derive our DQNM expansion formulas from Eq.~\eqref{PQNM}, the degree of the polynomial $g_\sigma(\lambda)$ can be set to be larger than 1 (as demonstrated by the bottom-right subfigure of Fig.~\ref{fig:e1}). In particular, the more poles the permittivity model has, the higher the degree of the polynomial $g_\sigma(\lambda)$ can be.

\subsection{Multi-pole model of permittivity}
\begin{figure}[ht!]
\centering\includegraphics[width=0.7\linewidth]{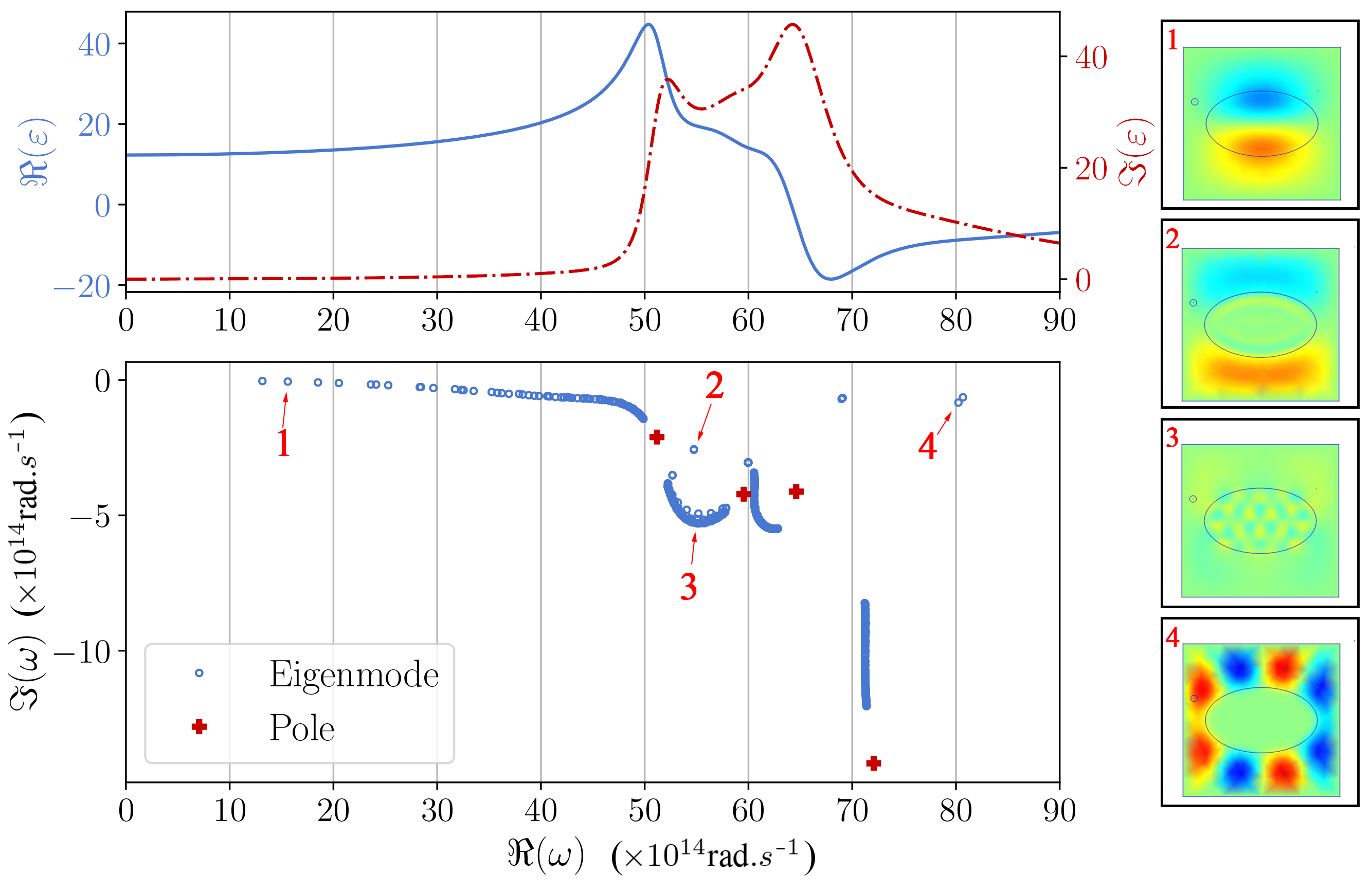}
\caption{Spectrum of complex eigenfrequencies (bottom left) corresponding to the 4-pole permittivity (top left). Four eigenfields (real part) are depicted at the right.}
\label{fig:eigen2}
\end{figure}

\begin{figure}[ht!]
\centering\includegraphics[width=0.72\linewidth]{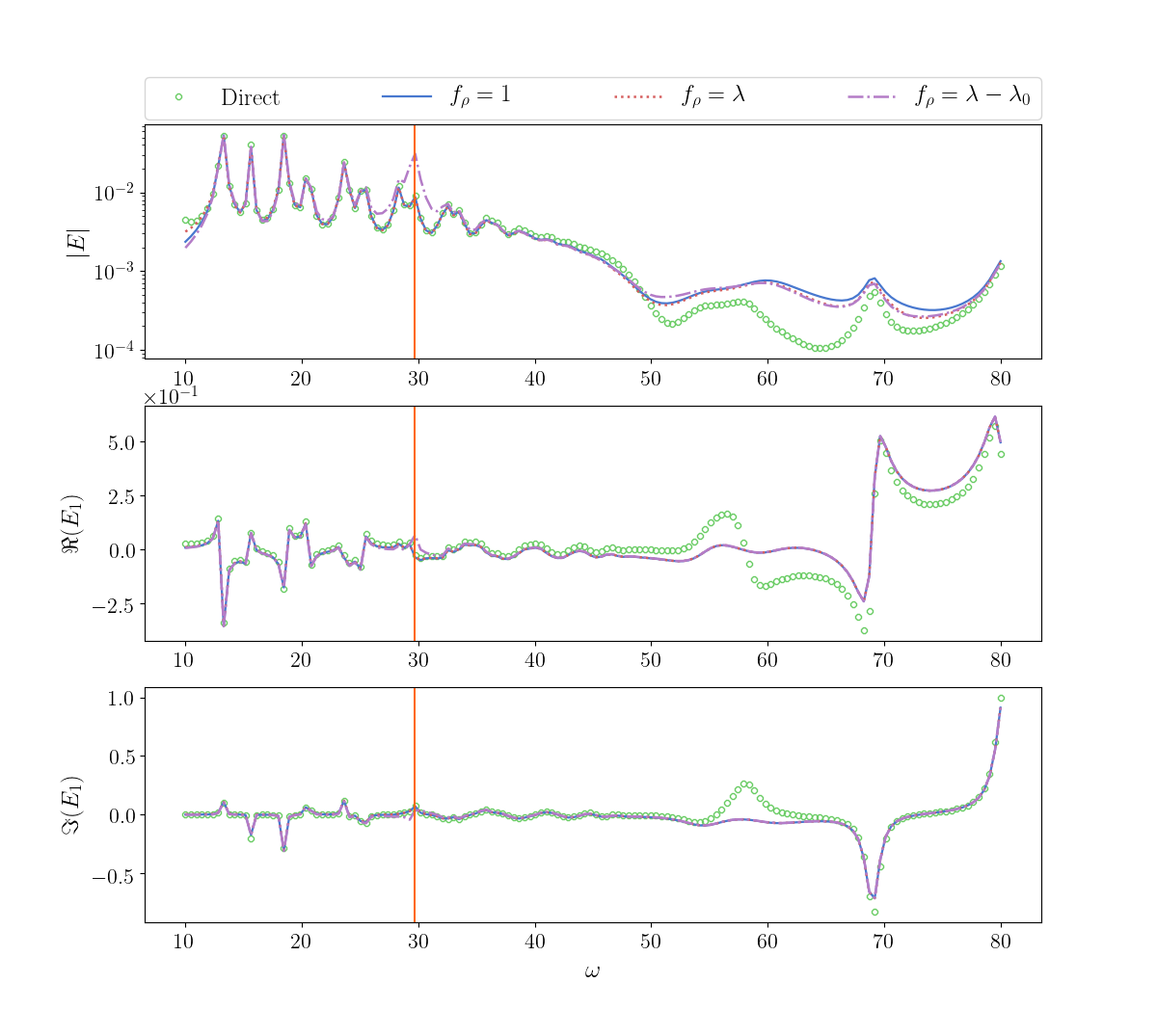}
\caption{Scattered field $\mathbf{E}$ obtained by expansion for different functions of $f_\rho$ (blue, red and purple curves) or by solving a direct problem classically (green dots) corresponding to the 4-pole permittivity. The orange vertical line indicates the value $\Re(\omega_0)$.}
\label{fig:e2}
\end{figure}

This subsection is intended to illustrate the difficulty when increasing the number of poles in the permittivity model Eq.~\eqref{pole}. The spectrum of complex resonances with the 4-pole permittivity model $N_p =4$ is shown in Fig.~\ref{fig:eigen2}. We especially emphasize the 4 poles (red crosses) which divide the complex plane in several sub-regions. In those sub-regions, we can distinguish two types of eigenfrequencies: Some are distributed separately (for example mode 1, 2, and 4 in Fig.~\ref{fig:eigen2}) which are responsible for the physical properties of the system; some gather into separate clusters (for example mode 3 in Fig.~\ref{fig:eigen2}) whose eigenfields oscillate with very high spatial frequencies. These `cluster' modes, which result from the accumulation points around the poles of the permittivity, affects the performance of the DQNM expansion around the corresponding frequencies (Fig.~\ref{fig:e2}). Indeed, the expansion shows good agreement with respect to the direct data (green dots) except around the frequencies of the `cluster' modes and the permittivity poles. It is also worth noting that when $f_\rho(\lambda)=\lambda-\lambda_0$, the expansion Eq.~\eqref{elec} endures discrepancies around $\Re(\omega_0)$ where $\omega_0=29.633-0.295i$ $(\times 10^{14} \textrm{rad}.s^{-1})$.

To sum up, by raising the number of poles, the permittivity model fits better with realistic materials. On the other hand, a large number of poles will fragment the complex plane of eigenfrequencies and limit the effective frequency range of our DQNM expansion.

\subsection{Magnetic fields and Plasmonic resonances}
The `cluster' modes around the poles of the permittivity model are not the only problem regarding the numerical implementation of DQNM expansion. In this subsection, we will discover the existence of the plasmonic resonances and their effects on DQNM expansion in the case of magnetic fields in TM polarization.

\begin{figure}[ht!]
\centering\includegraphics[width=0.88\linewidth]{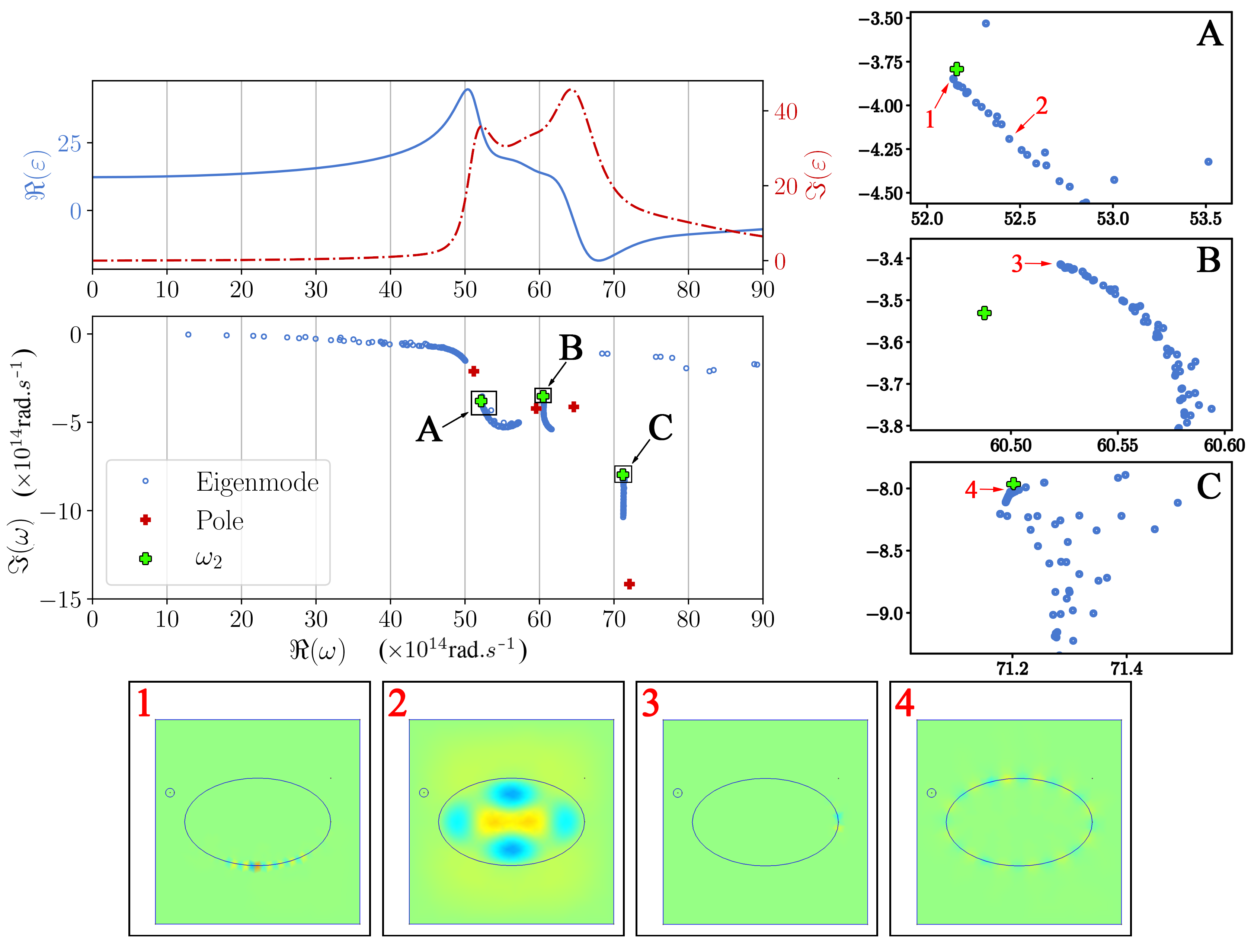}
\caption{Spectrum of complex eigenfrequencies (bottom left) of magnetic field  $\mathbf{H}$ corresponding to the 4-pole permittivity (top left). The green crosses refer to the plasmons, solutions of  $\varepsilon(\omega_2)=-1$ (There should be 4 green crosses but the forth one is out of our domain of interest). Four eigenfields (real part) are also depicted at the bottom.}
\label{fig:eigen3}
\end{figure}

\begin{figure}[ht!]
\centering\includegraphics[width=0.70\linewidth]{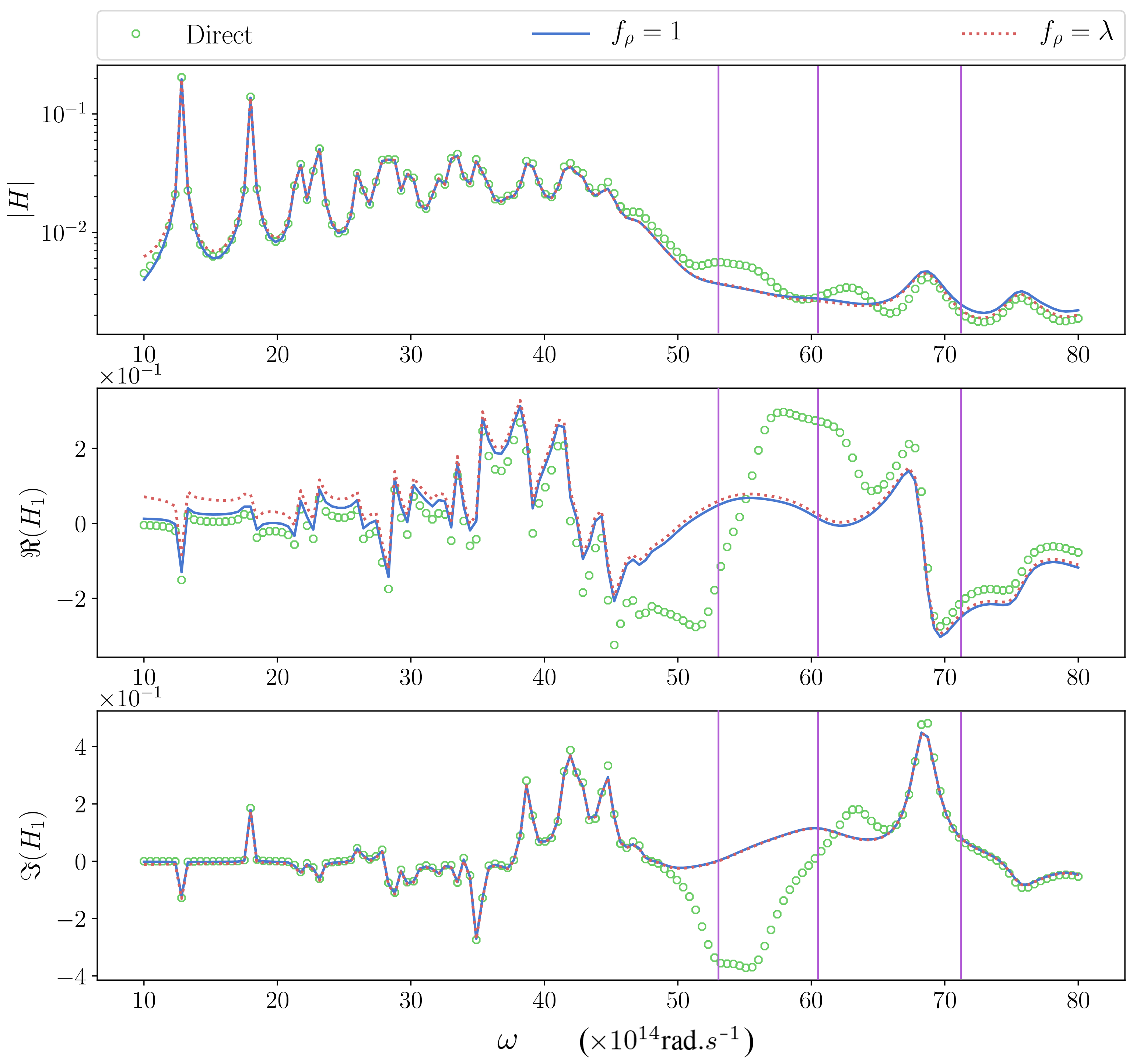}
\caption{Scattered field $\mathbf{H}$ obtained by expansion for different functions $f_\rho=1$ (blue curves) and $f_\rho=\lambda$ (red curves) or by solving a direct problem classically (green dots) corresponding to the 4-pole permittivity. The purple vertical lines indicate the positions of $\Re(\omega_2)$.}
\label{fig:h1}
\end{figure}

In particular, for the spectrum of eigenfrequencies of magnetic fields $\mathbf{H}$, there exists the second kind of accumulation points (see sidebar A, B, and C in Fig.~\ref{fig:eigen3} for more details), which locate around the plasmon branch where $\varepsilon(\omega_2)=-1$ (green crosses in Fig.~\ref{fig:eigen3}). These modes are indeed plasmonic resonances which distribute on the interface between $\Omega_0$ and $\Omega_1$ with the spatial frequencies tending to infinity (for example mode 1, 3, and 4 in Fig.~\ref{fig:eigen3}). They must be distinguished from the `cluster' modes caused by permittivity poles (mode 2 in Fig.~\ref{fig:eigen3}).  It is worth pointing out that the locations of these plasmonic resonances do not conspicuously converge (see inset C in Fig.~\ref{fig:eigen3}) to the analytical position $\omega_2$ where $\varepsilon(\omega_2)=-1$. In the case of polygonal sign-changing interfaces, the use of structured symmetric mesh stabilizes the numerical discretization of the plasmonic accumulation point \cite{demsy,mesh1}. However, the symmetry requirements with respect to a curved boundary remain to be elucidated. Unsurprisingly, the plasmonic accumulation points exacerbate the error of the DQNM expansion Eq.~\eqref{magn} in the frequency range nearby them (see Fig.~\ref{fig:h1}).


\subsection{Open structure and Perfectly Matched Layer (PML)}
In the previous computations, we have demonstrated the efficiency of the expansion formalisms in bounded structures. In practice, the electromagnetic system is often open, which makes the leaky resonant modes grow exponentially in space at infinity \cite{Sammut:76}. A solution is to use the Perfectly Matched Layers (PML) to truncate and damp the electromagnetic fields in free space \cite{PML}. Thus, in the last numerical example of this paper, we will mention the open structure as well as the effect of PML modes on the DQNM expansion.

\begin{figure}[ht!]
\centering\includegraphics[width=0.56\linewidth]{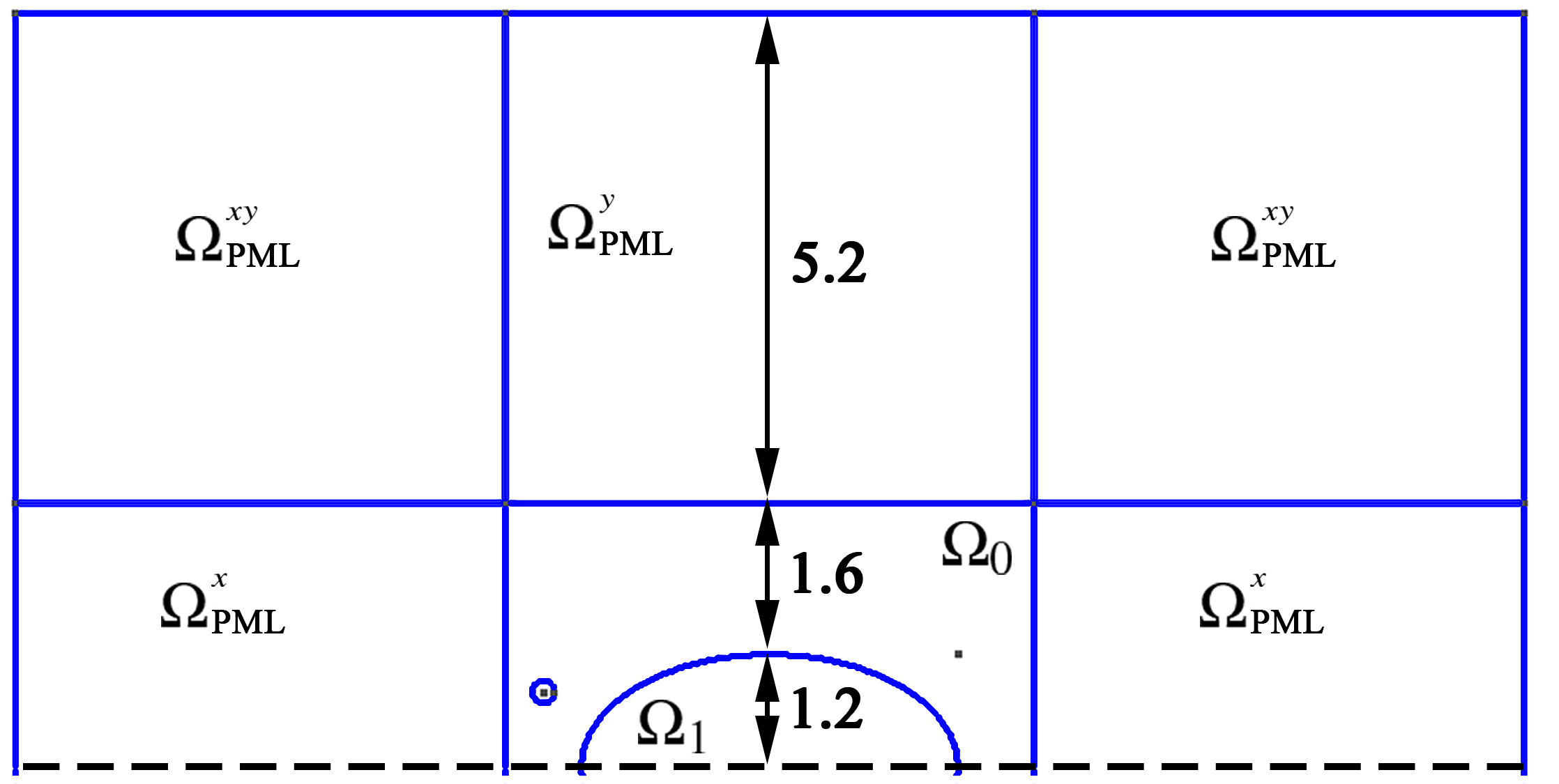}
\caption{The upper half of the geometry for the unbounded structure.}
\label{fig:PML}
\end{figure}

The PML layer $\Omega_{\textrm{PML}}$ is introduced as seen in Fig.~\ref{fig:PML}. We follow \cite{Vial2} in replacing the initial material properties $\pmb{\varepsilon}$ and $\pmb{\mu}$ in the PML domain $\Omega_{\textrm{PML}}$ (vacuum in this case) by equivalent material $\pmb{\varepsilon}^s$ and $\pmb{\mu}^s$ given by the following rule:
\begin{align*}
\pmb{\delta}_s \coloneqq \mathbf{J}_s^{-1} \pmb{\delta} \mathbf{J}_s^{-\intercal} \text{det}(\mathbf{J}_s) \quad \textrm{for} \quad \pmb{\delta} = \lbrace \pmb{\varepsilon},\pmb{\mu} \rbrace ,
\end{align*}
where $\mathbf{J}_s$ is the stretched Jacobian matrix such that:
\begin{align*}
 \mathbf{J}_s = \left\{  \begin{array}{ccc}
\textrm{diag}(s_x,1,1) & \textrm{in} & \Omega^x_\textrm{PML}\\  \textrm{diag}(1,s_y,1) & \textrm{in} & \Omega^y_\textrm{PML} \\ \textrm{diag}(s_x,s_y,1) & \textrm{in} & \Omega^{xy}_\textrm{PML}\\
\end{array} \right.,    
\end{align*}
where $s_x = s_y = \sigma \exp(i\phi)$ with $\sigma=1$ and $\phi =\pi/10$. The regions $\Omega^x_\textrm{PML}, \Omega^y_\textrm{PML}$, and $\Omega^{xy}_\textrm{PML}$ are depicted in Fig.~\ref{fig:PML}.

\begin{figure}[ht!]
\centering\includegraphics[width=0.75\linewidth]{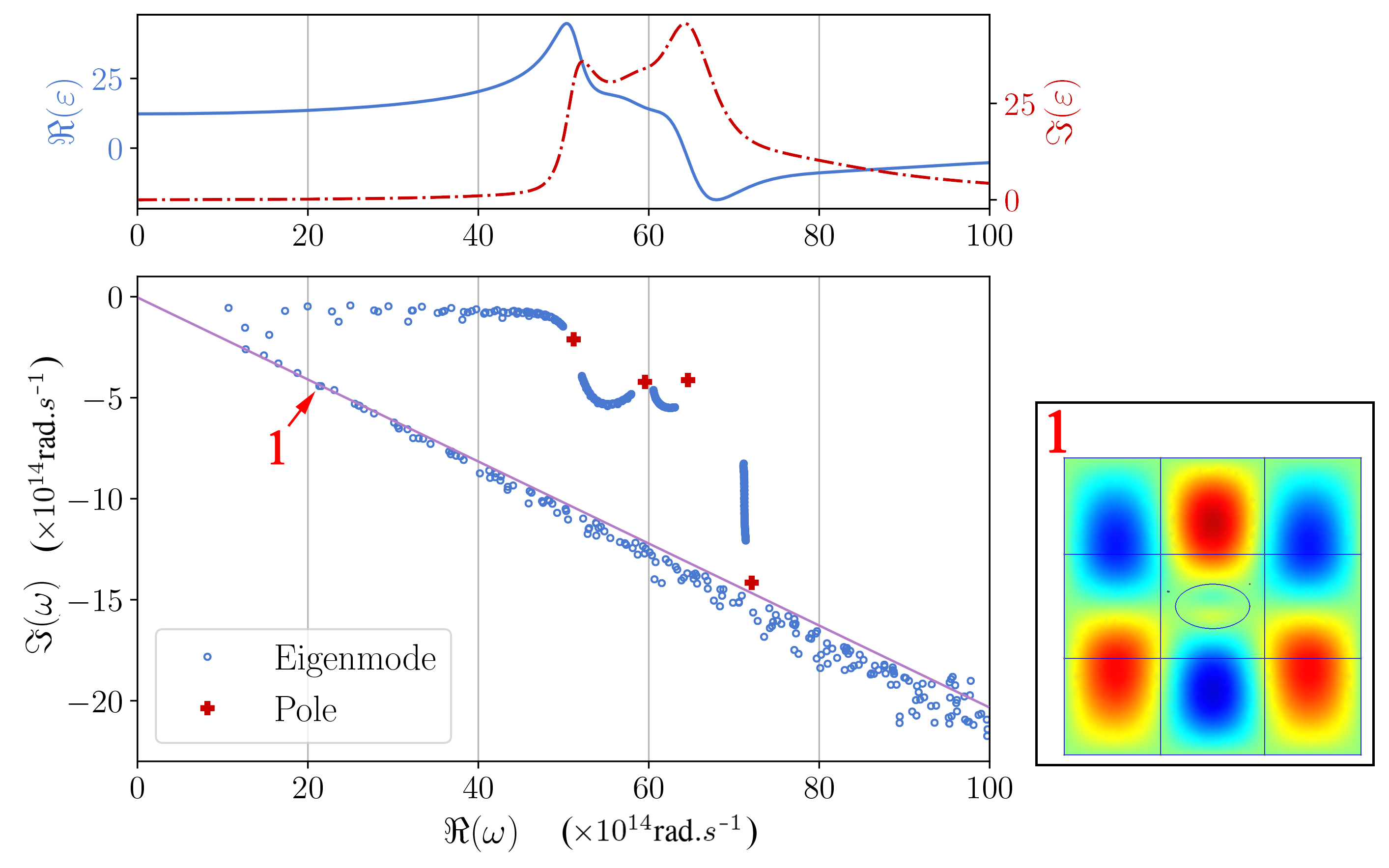}
\caption{Spectrum of complex eigenfrequencies (bottom left) of electric field in the unbounded structure. The eigenfield of PML mode is depicted on the right. The purple line illustrates the slope $\theta \approx -0.20456$.}
\label{fig:ei_4}
\end{figure}

We solve the eigenvalue problem of the electric field in TE polarization and the position of eigenfrequencies is shown in Fig.~\ref{fig:ei_4}. It is easy to see that the theoretical continuous spectrum, which is supposed to be located on $\mathbb{R}^+$, is rotated of an angle $\theta = -\text{arg}(2.8+5.2\exp(i\phi))\approx -0.20456 \,\text{rad}$ (For detailed instructions of calculation of the angle $\theta$, we refer the reader to \cite{NGUYEN}). This results in a large number of so-called discretized B\'erenger
`PML' modes \cite{PML,spectral}, whose eigenfield is concentrated in the PML region $\Omega_\textrm{PML}$ and far away from the scatterers (as can be seen from the field map of mode 1 in Fig.~\ref{fig:ei_4}). It should be noticed that the neatly aligned points corresponding to these modes are clearly becoming numerically unstable further on the curve as explained by the pseudo-spectrum theory of L. Trefethen
\cite{Trefethen}. 

\begin{figure}[ht!]
\centering\includegraphics[width=0.6\linewidth]{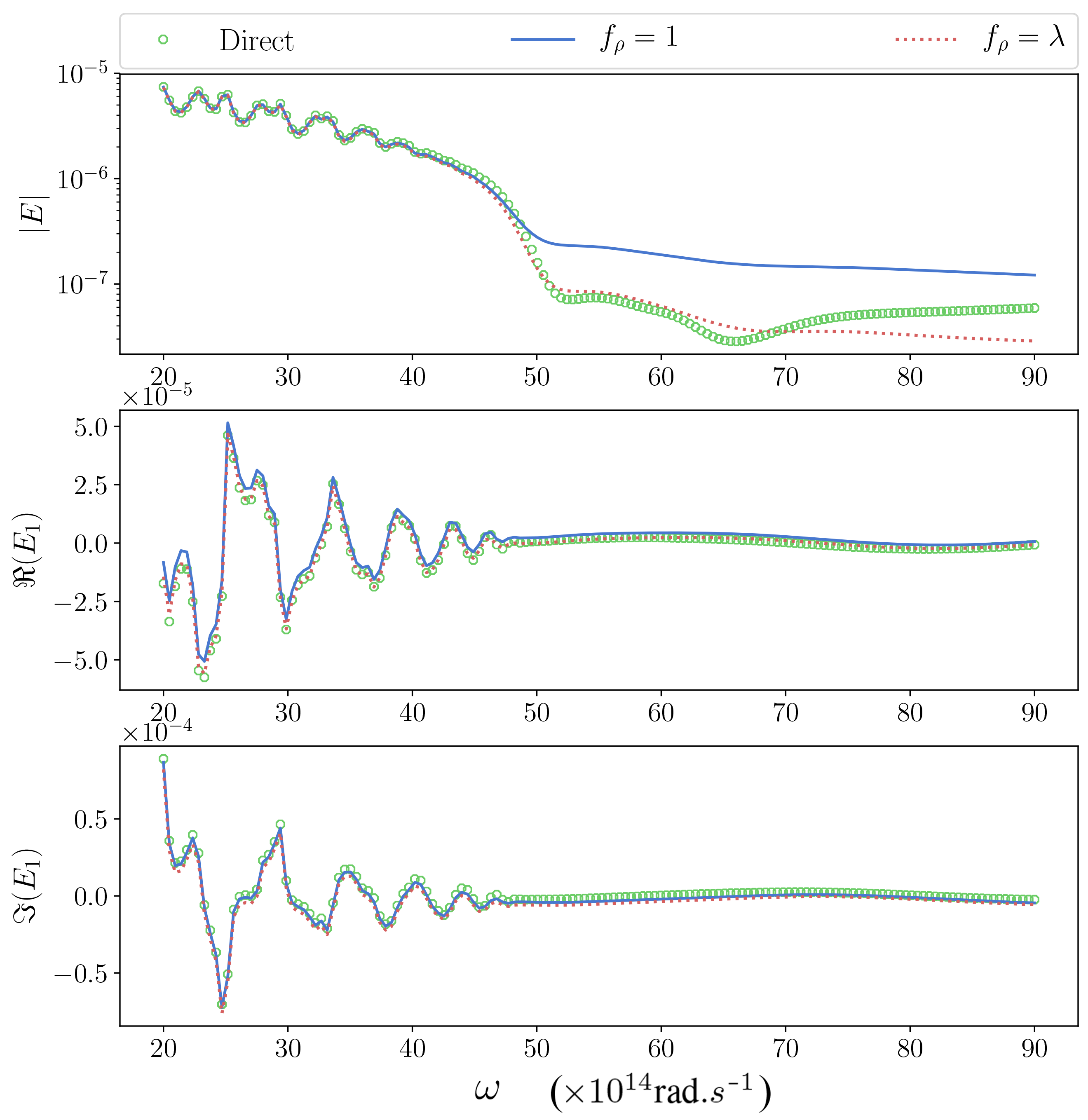}
\caption{Electric field obtained by expansion for different functions $f_\rho =1$ (blue curves) and $f_\rho =\lambda$(red curves) or by solving a direct problem classically (green dots) for the unbounded structure.}
\label{fig:e_4_1}
\end{figure}

We can see that the optical properties of the given open structure are fully captured by our DQNM expansion technique at low frequencies (see Fig.~\ref{fig:e_4_1}). When the frequency is larger, the discrepancies of the norm of electric fields inside the scatterer become noticeable since there is almost no field in the domain $\Omega_1$ (Pay attention that the figure is drawn in logarithmic scale and the values are really low). Numerical experiences show that the instability of PML modes (at high frequencies) may add noise to the DQNM expansion. In addition, we notice that the expansion with $f_\rho=\lambda$ provides a better approximation of the field inside the scatterer comparing to the case $f_\rho=1$.

\section{Conclusion}
In this paper, the non-uniqueness of the DQNM expansion formalisms is tackled: We systematically discussed the existence of a family of expansion equations, which are determined by the frequency-dependent factor $\frac{f_\rho(\lambda_n)}{f_\rho(\lambda)}$. By modifying this factor, we have been able to recover our previous result in \cite{Zolla} as well as other expansion formulas proposed in \cite{Sau}. Three examples of expansion formulas of the family are verified by numerical computation. They suggest that the function $f_\rho(\lambda)$ should be chosen in such a way that its roots do not overlap with our domain of interest. Moreover, we also demonstrate the possibility to use the multi-pole model of permittivity to impose the dispersion of realistic materials in the calculation. Even with great potential in practice, we still have to be extremely careful when applying modal expansion with this multi-pole model, since some discrepancies may occur in some particular frequency ranges:
\begin{itemize}
    \item In the vicinity of `cluster' modes, as the consequence of poles of the permittivity model.
    \item In the vicinity of the plasmonic branch where $\varepsilon(\omega_2)=-1$.
    \item At high frequencies due to the numerical instability of PML modes as explained by the pseudo-spectrum theory.
\end{itemize}

Finding a way to overcome these mentioned difficulties will guide the next development of the research on DQNM expansion. Meanwhile, the work will be extended to a wider range of materials as well as 3-D structures in the future.

\section*{Funding}
This research was supported by ANR RESONANCE project, grant ANR-16-CE24-0013 of the French Agence Nationale de la Recherche.

\section*{Disclosures}
The authors declare no conflicts of interest.

\section*{Appendix - Left eigenvalue problem}
\label{sec:app}
Solving the `left' eigenvalue problem is equivalent to find the eigensolutions of the adjoint operator:  $\mathcal{M}_{\pmb{\xi},\pmb{\chi}}^\dagger(\lambda_n) | \mathbf{w}_n \rangle = 0$. The adjoint operator of 
$\mathcal{M}_{\pmb{\xi},\pmb{\chi}}(\lambda)$, namely the operator $\mathcal{M}^{\dagger}_{\pmb{\xi},\pmb{\chi}}(\lambda)$ is defined by $\langle \mathbf{y}, \mathcal{M}_{\pmb{\xi},\pmb{\chi}}(\lambda) \mathbf{x}\rangle = \langle \mathcal{M}_{\pmb{\xi},\pmb{\chi}}^\dagger(\lambda) \mathbf{y},  \mathbf{x}\rangle $ for all $\mathbf{x},  \mathbf{y}$. For so doing, an integration by parts is necessary:
\begin{align}\label{adjoint}
    \langle \mathbf{y},\mathcal{M}_{\pmb{\xi},\pmb{\chi}}(\lambda)\mathbf{x} \rangle  = & \int_\Omega \overline{\mathbf{y}} \cdot \left(\nabla \times \left(\pmb{\xi}^{-1}(\lambda) \nabla \times \mathbf{x}  \right) + \lambda^2 \pmb{\chi}(\lambda) \mathbf{x} \right) \,d\Omega \nonumber \\ 
     = & \int_\Omega \left( \nabla \times \left( \pmb{\xi}^{-\intercal}(\lambda) \nabla \times \overline{ \mathbf{y}}\right) + \lambda^2 \pmb{\chi}^\intercal(\lambda) \overline{ \mathbf{y}} \right) \cdot \mathbf{x} \, d\Omega  \nonumber\\
    & + \int_{\partial\Omega} \left[ (\pmb{\xi}^{-\intercal}(\lambda)\nabla \times \overline{ \mathbf{y}}
) \cdot  (\mathbf{n}\times \mathbf{x} ) - (\mathbf{n}\times \overline{ \mathbf{y}}) \cdot (\pmb{\xi}^{-1}(\lambda) \nabla \times \mathbf{x}) \right]\, dS 
\end{align}
where the superscript $^{-\intercal}$ stands for the transposition of the inverse matrix: $\pmb{\xi}^{-\intercal} \coloneqq \left(\pmb{\xi}^{\intercal}\right)^{-1}=\left(\pmb{\xi}^{-1}\right)^{\intercal}$. The previous equation is obtained using $\int_\Omega \overline{\mathbf{y}}\cdot (\nabla \times \mathbf{x}) \, d\Omega= \int_\Omega (\nabla \times  \overline{\mathbf{y}}) \cdot \mathbf{x} \, d\Omega - \int_{\partial\Omega} (\overline{\mathbf{y}} \times \mathbf{x})\cdot \mathbf{n} \, dS $ and $\overline{\mathbf{y}}\cdot (\pmb{\xi}\mathbf{x}) = (\pmb{\xi}^\intercal\overline{\mathbf{y}}) \cdot \mathbf{x}$.

Under particular boundary conditions, for example homogeneous Dirichlet boundary condition $(\mathbf{n}\times \mathbf{x})\vert_{\partial\Omega}=0$ or homogeneous Neumann boundary condition $\mathbf{n}\times (\pmb{\xi}^{-1}\nabla \times  \mathbf{x})\vert_{\partial\Omega}=0$, the boundary term in Eq.~\eqref{adjoint} vanishes. Then, we have:
\begin{align*}
    \langle \mathbf{y},\mathcal{M}_{\pmb{\xi},\pmb{\chi}}(\lambda)\mathbf{x} \rangle 
     = \int_\Omega \left( \overline{ \nabla \times \left( \pmb{\xi}^{-\ast}(\lambda) \nabla \times \overline{ \mathbf{y}}\right) + \overline{\lambda}^2 \pmb{\chi}^\ast(\lambda) \overline{ \mathbf{y}}} \right) \cdot \mathbf{x} \, d\Omega 
   =  \langle \mathcal{M}_{\pmb{\xi},\pmb{\chi}}^\dagger(\lambda) \mathbf{y},\mathbf{x} \rangle 
\end{align*}
where the superscript $^{-\ast}$ denotes the conjugate transpose of the inverse matrix: $\pmb{\xi}^{-\ast} \coloneqq \left(\pmb{\xi}^{\ast}\right)^{-1}=\left(\pmb{\xi}^{-1}\right)^{\ast}$. In addition, due to the Hermitian symmetry of the permittivity and permeability, $\overline{\pmb{\xi}(\lambda)} = \pmb{\xi}(\overline{\lambda})$ and $\overline{\pmb{\chi}(\lambda)} = \pmb{\chi}(\overline{\lambda})$ , and, finally, the explicit form of the adjoint operator is expressed as follows:
\begin{align}\label{eq7}
    \mathcal{M}_{\pmb{\xi},\pmb{\chi}}^\dagger(\lambda) = \nabla \times \left( \pmb{\xi}^{-\intercal}(\overline{\lambda}) \nabla \times \cdot \right) +  \overline{\lambda}^2 \pmb{\chi}^\intercal(\overline{\lambda}) = \mathcal{M}_{\pmb{\xi}^\intercal,\pmb{\chi}^\intercal}(\overline{\lambda}).
\end{align}

If the material is reciprocal, \emph{i.e.} the permittivity and permeability are represented by symmetric tensors $\pmb{\xi} = \pmb{\xi}^\intercal$, $\pmb{\chi} = \pmb{\chi}^\intercal$, we have:
\begin{align}
   \overline{\mathcal{M}_{\pmb{\xi},\pmb{\chi}}^\dagger(\lambda_n) \mathbf{w}_n} = \nabla \times \left( \pmb{\xi}^{-1}(\lambda_n) \nabla \times \overline{\mathbf{w}_n} \right) +  \lambda^2 \pmb{\chi}(\lambda_n) \overline{\mathbf{w}_n} =0,
\end{align}
which implies the `left' eigenvectors $\mathbf{w}_n$ are the complex conjugate of their `right' counterpart, namely, $\mathbf{w}_n = \overline{\mathbf{v}_n}$.

It is worth noting that the complex conjugate relation between the `left' and `right' eigenvectors does not hold for all the geometric structures and boundary conditions. Indeed, this relation fails in the case of the diffraction grating computed with Bloch-Floquet quasiperiodicity conditions. The boundary then contains two parallel lines translated by a vector $\mathbf{d}$ where the fields at corresponding points are equal up to an $\exp(i\pmb{\kappa}\cdot\mathbf{d})$ phase factor, the $\pmb{\kappa}$ being a given reciprocal space vector. In this case, both the `left' and `right' eigenvectors must be Bloch waves:
\begin{align*}
\mathbf{w}^{\pmb{\kappa}}_n(\mathbf{r}) = \mathbf{w}^{\pmb{\kappa}}_{\#n}(\mathbf{r}) \exp(i\pmb{\kappa}\cdot \mathbf{r}) \qquad
    \mathbf{v}^{\pmb{\kappa}}_n(\mathbf{r}) = \mathbf{v}^{\pmb{\kappa}}_{\#n}(\mathbf{r}) \exp(i\pmb{\kappa}\cdot \mathbf{r})
\end{align*}
where $\mathbf{w}^{\pmb{\kappa}}_{\#n}$ and $\mathbf{v}^{\pmb{\kappa}}_{\#n}$ are $\mathbf{d}$-periodic functions 
(see \cite{duy1,duy2,vial} for more details).

Then, it is clear that $
    \overline{\mathbf{w}^{\pmb{\kappa}}_n(\mathbf{r})} = \overline{\mathbf{w}^{\pmb{\kappa}}_{\#n}(\mathbf{r})} \exp(-i\pmb{\kappa}\cdot \mathbf{r})$,
which implies that the relation $\mathbf{w}_n = \overline{\mathbf{v}_n}$ is no longer true in this case. In particular, the equality $\mathbf{w}^{\pmb{\kappa}}_n=\overline{\mathbf{v}^{\pmb{\kappa}}_n}$ only holds when the dephasing term $\exp(i\pmb{\kappa}\cdot \mathbf{d})$ equals to 1, \emph{i.e.} $\pmb{\kappa}=\mathbf{0}$. In other cases, \emph{i.e.} when dealing with non-reciprocal materials or with diffraction gratings in non-normal incidence, the notion of eigentriplets becomes crucial because there are no longer simple link between the 
`bra' and the `ket' eigenvectors.

\bibliography{sample}

\end{document}